\documentclass[11pt,a4paper]{article}
\pdfoutput=1

\usepackage{jcappub} 
\usepackage{wasysym}
\usepackage{enumitem}
\usepackage{color}

\newcommand{\mpl}{M_{P}}

\newcommand{\GeV}{{\rm GeV}}
\newcommand{\Mpc}{{\rm Mpc}}
\newcommand{\td}{{\rm d}}

\newcommand{\msol}{M_{\astrosun}}
\newcommand{\rpbh}{\rho_{\rm PBH}}
\newcommand{\rtot}{\rho_{\rm tot}}
\newcommand{\prs}{\mathcal{P}_{\zeta}}

\title{Single Field Double Inflation and Primordial Black Holes}

\author[a]{K. Kannike,}
\author[a,b]{L. Marzola,}
\author[a]{M. Raidal,}
\author[a]{and H. Veerm\"ae}

\affiliation[b]{ Institute of Physics, University of Tartu, W. Ostwaldi 1,  50411 Tartu, Estonia}
\affiliation[a]{National Institute of Chemical Physics and Biophysics, R\"avala 10, 10143 Tallinn, Estonia}

\vspace*{1cm}

\emailAdd{kristjan.kannike@cern.ch}
\emailAdd{luca.marzola@cern.ch}
\emailAdd{martti.raidal@cern.ch}
\emailAdd{hardi.veermae@cern.ch}

\vspace{3cm}

\abstract{Within the framework of scalar-tensor theories, we study the conditions that allow single field inflation dynamics on small cosmological scales to significantly differ from that of the large scales probed by the observations of cosmic microwave background. The resulting single field double inflation scenario is characterised by two consequent inflation eras, usually separated by a period where the slow-roll approximation fails. At large field values the dynamics of the inflaton is dominated by the interplay between its non-minimal coupling to gravity and the radiative corrections to the inflaton self-coupling. For small field values the potential is, instead, dominated by a polynomial that results in a hilltop inflation. Without relying on the slow-roll approximation, which is invalidated by the appearance of the intermediate stage, we propose a concrete model that matches the current measurements of inflationary observables and employs the freedom granted by the framework on small cosmological scales to give rise to a sizeable population of primordial black holes generated by large curvature fluctuations. We find that these features generally require a potential with a local minimum. We show that the associated primordial black hole mass function is only approximately lognormal.}


\date{\today}

\begin{document}
\maketitle
\flushbottom

\section{Introduction} 
\label{sec:Introduction}

The observation of coalescence of ${\mathcal O}(10)$ solar mass black holes by the LIGO~\citep{Abbott:2016blz,Abbott:2016nmj} collaboration has revived interest in primordial black holes (PBHs)~\citep{Hawking:1971aa,Carr:1974nx,1975ApJ...201....1C,1975A&A....38....5M,CHAPLINE:1975aa}, which were earlier investigated as a possible solution to the puzzle of the origin of dark matter (DM). PBH DM may have been created by large density perturbations, $\delta\rho/\rho\sim 1$, that were produced during the last stages of inflation~\citep{Carr:1994ar,GarciaBellido:1996qt,Bullock:1996at,Kawasaki:1997ju,Yokoyama:1998pt,Kawasaki:1998vx,Kohri:2007qn,Kawaguchi:2007fz,Frampton:2010sw,Drees:2011hb,Drees:2011yz,2011arXiv1107.1681L, Bugaev:2011wy,Kawasaki:2012wr,Kohri:2012yw,Linde:2012bt,Bugaev:2013vba,Bugaev:2013fya,Clesse:2015wea,Kawasaki:2016pql,Garcia-Bellido:2016dkw,Inomata:2016rbd,Orlofsky:2016vbd,Garcia-Bellido:2017mdw,Inomata:2017okj,Domcke:2017fix,Ezquiaga:2017fvi,Belotsky:2014kca,Cheng:2016qzb}. Taking into account the most recent constraints,\footnote{Constraints for realistic, non-monochromatic PBH mass distributions taking into account most recent updates are derived in Ref.~\citep{Carr:2017jsz}. For earlier reviews see~\citep{Carr:2009jm,Carr:2016drx}.} PBHs may still constitute a large fraction of the measured DM abundance~\citep{Bird:2016dcv,Clesse:2016vqa,Sasaki:2016jop,Blinnikov:2016bxu}.
Such a DM component could help explaining the rapidity of structure formation at small scales and provide, on top of that, the seeds for supermassive black holes and therefore for galaxy formation~\cite{Khlopov:2004sc,Mack:2006gz,Rubin:2001yw,Kawasaki:2012kn,Clesse:2015wea}. The properties of PBHs could also explain other astrophysical and cosmological puzzles (for a review and references see~\citep{Garcia-Bellido:2017fdg}).

All the current measurements of the cosmic microwave background (CMB) perturbations~\citep{Hinshaw:2012aka,Ade:2015xua} are in good agreement with the predictions of single field inflation. Observations imply a decreasing, red-tilted, power spectrum, as well as presently unobservable non-Gaussianity and isocurvature effects~\citep{Ade:2015lrj} that strongly constrain possible inflationary potentials.In order to comply with the Planck measurements and allow for the possible existence of large perturbations at the end of inflation that source the PBH production, the inflationary era must comprise two very different stages. The first phase of inflation must create the large scale fluctuations observed in the CMB, while the second phase of inflation is responsible for the generation of large fluctuations at small cosmological scales. Therefore, the most natural framework to realise such a two stage inflation involves multiple fields, such as in the hybrid inflation~\citep{GarciaBellido:1996qt,Kawasaki:1997ju,Kohri:2007qn,Polarski:1992dq}. However, single field scenarios can also accommodate the desired second phase provided that the inflaton crosses a near-inflection point in the part of its potential responsible for the last period of inflation~\citep{Bullock:1996at, Yokoyama:1998pt,Garcia-Bellido:2017mdw,Ezquiaga:2017fvi}. This scenario, the so-called single field double inflation, is already severely constrained by current measurements of scalar spectral index $n_s,$ tensor-to-scalar ratio $r$, amplitude of the power spectrum $A_{s}$ at the pivot scale $k_{*} = 0.05 \Mpc^{-1}$, as well as by the number of $e$-folds $N$. Nevertheless, viable prototypal single field double inflation models were recently proposed in Ref.~\citep{Garcia-Bellido:2017mdw,Ezquiaga:2017fvi}, where the main features required of the single field inflationary potential were considered. For completeness we remark that in certain models of single field inflation with a singular potential~\cite{Starobinsky:1992ts} it is also possible to produce the observed DM abundance in PBH~\cite{Blais:2002nd}.

The aim of the present paper is to develop a consistent theory of single field double inflation based on exact classical dynamics of the scalar field. We propose a realistic, particle physics motivated and UV complete scenario of single field double inflation that could explain simultaneously the Planck measurements of CMB fluctuations and create a sizeable population of PHBs at the end of inflation. To be as general as possible, we consider single scalar inflation in the framework of scalar-tensor theories, that, in addition to a general potential, allow for a modified kinetic term and a non-minimal coupling to gravity. We require the inflaton potential to be renormalisable and UV complete and, therefore, we consider a tree-level Lagrangian containing operators of at most dimension four. In order to capture the full implications of the single field double inflation dynamics, we perform our analysis by solving numerically the exact equation of motion for the inflaton field rather than relying on the slow-roll approximation. We find, in fact, that single field double inflation models typically violate the slow-roll conditions between the two stages of inflation, and that, consequently, the slow-roll approximation leads to inaccurate predictions. We also find that a mere inflection point is not able to guarantee that the inflaton field is sufficiently slowed down for the onset of the second inflationary era.

The more careful mathematical treatment and the physical origin of the proposed potential distinguish our work from the scenarios previously considered in literature. In our study we also allow for possible renormalisation effects that induce logarithmic running of the inflaton self-coupling. The emerging scenario, that is consistent with the CMB measurements, comprises a first stage of inflation regulated by the interplay between the non-minimal coupling of the inflaton and the logarithmic corrections due to renormalization effects. Such potentials were considered before in Refs.~\citep{Marzola:2015xbh,Marzola:2016xgb} and give rise here to a first stage of inflation that lasts for about 30-40 $e$-folds and differs from the most popular scenarios of Higgs~\citep{Bezrukov:2007ep} or Starobinsky~\citep{Starobinsky:1980te} inflations.

The first inflationary stage is then followed by a second phase of hilltop inflation~\citep{Boubekeur:2005zm},  which lasts for about 20-30 $e$-folds and creates the large density perturbations that source the PBH production. To this purpose, we arrange the potential to have a local minimum at low field values that slows down the inflaton field and flattens the potential enough to create a peak in the curvature fluctuation power spectrum. We stress again that a simple inflection point does not serve the purpose despite what often is stated in the literature: precise computations beyond the slow-roll approximation show that the required potential must be more complicated. As a consequence, the predicted PBH mass function may deviate from the lognormal shape that is typically considered in the literature~\citep{Green:2016xgy,Horowitz:2016lib,Kuhnel:2017pwq,Carr:2017jsz}  and have an extended low mass tail. We find that the amount of PHBs created during the second phase of inflation can easily saturate the present  bound on the DM abundance.

The paper is organised as follows. In section~\ref{sec:Inflation} we present the general framework of single field double inflation and describe the qualitative features of both the phases of inflation. In section~\ref{sec:model}  we present a concrete realisation of the results of section~\ref{sec:Inflation}. Details of the second phase of the inflation and the PBH DM abundance and mass distribution are studied in section~\ref{sec:PBH}. We conclude in section~\ref{sec:Conclusions}.
  
\section{Single field double inflation} 
\label{sec:Inflation}

The double inflation scenario is realised with potentials that possess flat regions or local minima able to significantly slow down the inflaton field after the first period of inflation has ended, thereby triggering a new phase of inflation. We find that the slow-roll regime can be barely maintained between the two inflationary phases even for potentials that are sufficiently flat as, in this case, small corrections can induce large changes in the effective evolution of the field. In presence of a local minimum, instead, the slow-roll approximation inevitably fails because the kinetic energy of the inflaton field needs to be sufficiently large to overcome the potential barrier. In the following we will detail these possibilities after briefly reviewing the relevant features of the inflationary Universe. We note also that a transition regime where slow-roll is violated has also been studied in the context of two-field inflation~\cite{Polarski:1992dq}.

\subsection{General considerations}

Our framework is that of single field inflation, specified in absence of interactions between inflaton and matter by three arbitrary functions by~\citep{Jarv:2016sow}
\begin{equation}\label{eq:L_jordan}
	\mathcal{L} = \sqrt{-g} \left( -\frac{1}{2}\mpl^{2} \Omega(\sigma) R + \frac{1}{2}K(\sigma) (\partial \sigma)^{2} - V(\sigma) \right),
\end{equation}
where $\Omega(\sigma)$ contains the non-minimal coupling, $K(\sigma)$ accounts for the possibility of a non-canonical kinetic term, $V(\sigma)$ is the potential and $\mpl =  2.4 \times 10^{18}~\GeV$ is the reduced Planck mass. In the following we neglect the cosmological constant and set the potential to vanish at the absolute minimum. Two of the three functions can be absorbed by means of a conformal transformation and a redefinition of the scalar field. The conformal transformation $g_{\mu\nu} \to \Omega(\sigma)^{-1} g_{\mu\nu}$ is used to recast the Lagrangian in the Einstein frame
\begin{equation}\label{eq:L_einstein}
	\mathcal{L} = \sqrt{-g} \left( -\frac{1}{2}\mpl^{2} R + \frac{1}{2} \bar K(\sigma) (\partial \sigma)^{2} - \bar V(\sigma) \right),
\end{equation}
where
\begin{equation}
	\bar K =  \frac{K}{\Omega} + \frac{3 \mpl^{2}}{2}\left(\frac{\partial_{\sigma}\Omega}{\Omega} \right)^{2},	
	\qquad
	\bar V = \frac{V}{\Omega^{2}}.
\end{equation}
The canonical Einstein frame field $\phi$ is then found as a solution of $\td \phi = \sqrt{\bar K} \td \sigma$. Because the analytic expression for $\phi(\sigma)$ is generally quite complicated (it may not even be given in a closed form), in the following we will retain the non-canonical kinetic term and analyse the inflaton dynamics in the Einstein frame.

The evolution of the Friedmann-Robertson-Walker inflationary Universe is dictated by the Friedmann equation and the scalar field equation,
\begin{equation}\label{eq:sfeq,Feq}
	3 \mpl^{2} H^{2} = \rho_{\phi}, \qquad
	\ddot \phi+ 3H \dot{\phi} + \partial_{\phi}\bar{V} = 0,	
\end{equation}
where $\rho_{\phi} = \frac{1}{2} \dot \phi^{2} + \bar{V} $ is the energy density of the scalar field. In the following we express time in terms of the number of $e$-folds by using the relation $\partial_{t} = H \partial_{N}$. The equations \eqref{eq:sfeq,Feq} can then be combined into a single equation
\begin{equation}\label{eq:exact_phiN}
	\partial_{N}^{2} \phi+ \left(3 - \frac{(\partial_{N} \phi)^{2}}{2\mpl^{2}}  \right)\left( \partial_{N} \phi + \mpl^{2}\frac{\partial_{\phi}\bar{V} }{\bar{V} } \right)= 0,
\end{equation}
that can be expressed in terms of the non-canonical variable $\sigma$ through $\partial_{N}\phi = \sqrt{\bar{K}} \partial_{N}\sigma$.

The qualitative behaviour of the inflating Universe is characterised by the Hubble slow-roll parameter~\citep{Liddle:1994dx}
\begin{equation}\label{eq:eps_H}
	\epsilon_{H} \equiv -\frac{\partial_{N}  H}{H} = \frac{(\partial_{N} \phi)^{2}}{2\mpl^{2}},
\end{equation}
that is related to the equation of state parameter of the inflaton field by $w = -1 + 2/3\,\epsilon_{H} $. Thus, inflation occurs whenever $\epsilon_{H} < 1$. The zeroes of the factors in brackets in Eq.~\eqref{eq:exact_phiN} define two broad classes of solutions: the first, with $\epsilon_{H} \approx 3$, is the kination dominated regime in which the field rolls at maximal speed with respect to $N$. Because this solution is a repeller, the kination dominated regime has necessarily a finite duration. The second solution, characterised by $\partial_{N} \phi \approx - \mpl^{2}\partial_{\phi}\bar{V} /\bar{V} $, describes the slow-roll regime where any dependence on the initial velocity is rapidly damped by Hubble friction $\epsilon_{H}$, determined as a function of the field. It is then customary to define the potential slow-roll parameters
\begin{equation}\label{def:slowroll_params}
	\epsilon_{V} =  \frac{\mpl^{2}}{2} \left( \frac{\bar V'}{ \bar V} \right)^{2},
	\qquad
	\eta_{V} =  \mpl^{2}\frac{ \bar V''}{\bar V},
	\qquad
	\xi_{V} = \mpl^4 \frac{ \bar V' \bar V'''}{\bar V^{2}},
\end{equation}
where the prime denotes differentiation with respect to the Einstein frame field $\phi$. The corresponding expressions in terms of the Jordan frame field $\sigma$ are obtained by using $\partial_{\phi} = \bar K^{-1/2}\partial_{\sigma}$.  The vanishing of the second bracket in \eqref{eq:exact_phiN}, that is slow-roll, is equivalently expressed as $\epsilon_{H} \approx \epsilon_{V}$. For consistency, slow roll should be far from the kination dominated regime, that is $\epsilon_{H} \ll 3$, which in turn imposes the conditions $\epsilon_{V}, \eta_{V} \ll 1$ on the slow-roll potential parameters.

The inflationary observables can be expressed in terms of $\epsilon_{H}$ and its derivatives. The power spectrum of scalar curvature perturbations is given by
\begin{equation}\label{eq:PRk}
	\prs(k) =  \frac{1}{8 \pi^{2} \mpl^{2}} \frac{H^{2}}{\epsilon_{H}}\,,
\end{equation}
where the right hand side is evaluated at the moment when the scale $k$ exits the horizon: for $k = aH$. The measurement of the amplitude of the scalar power spectrum $A_{s} \equiv \prs(0.05 \,\Mpc^{-1})$ gives $\ln(10^{10}A_{s}) = 3.089 \pm 0.036$~\citep{Ade:2015lrj}. The tensor-to-scalar ratio $r$ and the tilt of the spectrum $n_{s}$ are
\begin{equation}\label{eq:SR_r_ns}
	r = 16 \epsilon_{H}, \qquad n_{s} = 1 - 2 \epsilon_{H} + \partial_{N}\ln\epsilon_{H},
\end{equation}
and their experimental values are given by $r \lesssim 0.1$ and $n_{s} = 0.9655 \pm 0.0062$~\citep{Ade:2015lrj}. The number of $e$-folds characterising the length of inflation follows from \eqref{eq:eps_H},
\begin{equation}\label{eq:Ntot}
	N 
	=  	\frac{1}{\mpl} \int^{\phi_{N}}_{\phi^{\rm end}} \frac{\td \phi}{\sqrt{2\epsilon_{H}}} 
	\approx	\frac{1}{\mpl^{2}} \int^{\sigma_{N}}_{\sigma^{\rm end}} \frac{\bar K\, \bar V}{\partial_{\sigma} \bar V} \td \sigma,
\end{equation}
where the field value $\sigma^{\rm end}$ is such that $\epsilon(\sigma^{\rm end}) = 1$. The above equation for the number of $e$-folds applies in the slow-roll approximation and is redundant within the exact approach based on the solutions of Eq.~\eqref{eq:exact_phiN}.

In our analysis we apply the exact classical equations for the scalar field, Eq.~\eqref{eq:exact_phiN}, while using the approximate predictions \eqref{eq:PRk} for the spectrum of curvature perturbations. Eq. \eqref{eq:PRk} is commonly derived by assuming the slow-roll approximation~\cite{Martin:2007bw}. However, studies that go beyond slow-roll indicate that the relative error is expected to be of order $\epsilon_{H}$~\cite{Schwarz:2001vv}. It is therefore plausible that our analysis of the power spectrum for the transition phase between the two inflationary epochs may be inaccurate because $\epsilon_{H} \ll 1$ generally does not hold  there. These discrepancies, however, do not affect our conclusions for the two main inflationary stages as they occur in a slow-roll regime.

\subsection{The first phase of inflation} 
\label{sec:infphase1}

The density perturbations generated during the first phase of inflation are constrained by the measurements of the cosmic microwave background. The general features of inflationary dynamics at this scale can be discussed in a model-independent way in terms of the large-$N$ formalism \citep{Boyanovsky:2005pw,Roest:2013fha,Garcia-Bellido:2014gna}. The total number of $e$-folds, usually estimated to lie in the interval 50-60~\citep{Ade:2015lrj}, can be split into
\begin{equation}
	N = N_{1} + N_{2},
\end{equation}
where $N_{1}$ and $N_{2}$ denote the lengths of the first and the second inflationary phase respectively. In the case of double inflation, the spectral features of the CMB depend solely on $N_{1}$, therefore observables can be given on the large-$N$ formalism in terms of expansions in $1/N_{1}$ rather than $1/N$. As the total number of $e$-folds is bounded from above, a long second phase implies a shorter first phase. 

In this paper we focus on a particle physics motivated model based on a tree-level Lagrangian that contains operators with a mass dimension of 4 at most. This implies that $\Omega$, $K$ and $V$ in the Jordan frame Lagrangian \eqref{eq:L_jordan} are polynomials with possible logarithmic dependencies that arise from quantum corrections. Our numerical results show that in that case it is possible to approximate
\begin{equation}\label{eq:f1_r_ns}
	r \approx \frac{c_{r}}{N_{1}^{p}}, \qquad 
	n_{s} \approx 1 - \frac{r}{8} - \frac{p}{N_{1}},
\end{equation}
where $c_{r}$ and $p$ are constants depending on the parameters of the model. For the class of models under study $1 \leq p \leq 2$. 

It follows from Eq.~\eqref{eq:f1_r_ns} that for $r \approx 0$ (supported by current data), $n_{s} - 1$ is inversely proportional to the number of $e$-folds $N_{1}$. For example, choosing $N_{1} = 30$ and $r = 0$, and imposing that $n_{s} > 0.95$ yields $p < 1.5$. As a result, the dynamics of inflation during the first stage will necessarily deviate from the prediction of Higgs and Starobinsky inflation ($p \approx 2$~\citep{Starobinsky:1983zz,Giudice:2014toa}), as well as hilltop inflation ($p>2$~\citep{Boubekeur:2005zm,Roest:2013fha}). Considering a larger number of total $e$-folds or a shorter second inflationary phase will relax the bounds on $N_{1}$ and on $p$.

For the above reasons, single field double inflation naturally favour models predicting an almost scale invariant spectrum with $n_{s} \approx 1$ and a small $r \approx 0$ if a usual length of inflation $N = 50-60$ is assumed. Concrete examples of models that can successfully match the measured inflationary observables by means of a reduced first stage of inflation are the minimally coupled inflaton with an asymptotically logarithmic potential~\citep{Dvali:1994ms} or a non-minimally coupled inflaton with a running quartic~\citep{Marzola:2016xgb}. Once a model with the suitable large $N$-behaviour is identified, our framework allows to combine it with a suitable completion that will determine the dynamics at smaller scales (or small field values) without notably altering the predictions of \eqref{eq:f1_r_ns} at large field values.

\subsection{The second phase of inflation} 
\label{sec:infphase2}

It follows from Eq.~\eqref{eq:PRk} that the power spectrum is enhanced for small $\epsilon_{H}$. To illustrate the underlying mechanism, we first the slow-roll approximation for which $\epsilon_{H} \approx \epsilon_{V}$ so it is possible to express the slow-roll parameter in terms of the field only. However, our numerical study of the exact field equation \eqref{eq:exact_phiN} shows that double inflation models where the slow-roll approximation applies at all times are rare and, as a rule, do not generate large peaks in the power spectrum. Basic features of two generic classes of single field double inflation models are discussed at the end of this section.

\subsubsection{Slow-roll approximation}

A peak in the power spectrum corresponds to a local minimum of $\epsilon_{V}$ at $\phi_{c}$, that is $\partial_{\phi}\epsilon_{V}(\phi_{c}) = 0$; in the extreme case, also $\epsilon_{V}(\phi_{c}) = 0$.  If $\epsilon_{V}(\phi_{c}) = 0$, under the slow-roll approximation Eq.~\eqref{eq:Ntot} then implies that the field stops at $\phi_{c}$ and gives rise to an ever-lasting period of inflation. More realistically, $\epsilon_{V}(\phi_{c})$ could be small but non-vanishing at $\phi_{c}$, in which case we can expand
\begin{equation}
	\epsilon_{V} = \epsilon_{V}(\phi_{c}) + \frac{1}{2} \epsilon''_{V}(\phi_{c}) (\phi - \phi_{c})^{2} + \mathcal{O}(\phi - \phi_{c})^{3}.
\end{equation}
Only small deviations from $\epsilon(\phi_{c}) = 0$ are considered, specifically we assume that $\epsilon_{V}(\phi_{c}) \ll  \mpl^{2}\epsilon''_{V}(\phi_{c})$. From the definition of the slow-roll parameters \eqref{def:slowroll_params} it follows that at the minimum, that is when $\epsilon'_{V}(\phi_{c}) = 0$, it can be shown that $\mpl^{2}\, \epsilon''_{V}(\phi_{c}) = \xi_{V}(\phi_{c}) - 4\epsilon^{2}_{V}(\phi_{c}) \approx \xi_{V}(\phi_{c})$, where the $\epsilon^{2}_{V}(\phi_{c})$ term is negligible by assumption. Expanding the square root in \eqref{eq:Ntot} yields an approximation of the length of the second phase of inflation. The number of $e$-folds that the inflaton spends around $\phi_{c}$,
\begin{equation}\label{eq:N2}
	N_{2} \approx  \frac{2\pi }{\sqrt{2 \xi_{V}(\phi_{c})}},
\end{equation}
 gives the length of the second phase of inflation. The maximal amplitude \eqref{eq:PRk} of scalar perturbations during this phase is
\begin{equation}\label{eq:PRc}
	\prs(k_{c}) =   \frac{1}{24 \pi^{2} \mpl^{4}} \frac{\bar V_{V}(\phi_{c})}{\epsilon_{V}(\phi_{c})},
\end{equation}
where $k_{c}$ corresponds to the horizon size when $\phi = \phi_{c}$. The shape of the amplitude \eqref{eq:PRk} of scalar perturbations generated during the second phase is well approximated by
\begin{equation}\label{eq:PRkc}
	\prs(k) 
	\approx  \prs(k_{c}) \left[1+ 2 \tan^{2}\left(\sqrt{\frac{\xi_{V}(\phi_{c})}{2}} \ln \frac{k}{k_{c}} \right)\right]^{-1},
\end{equation}
where we have used $k \propto e^{N}$. These results agree with Ref.~\citep{Garcia-Bellido:2017mdw}.

If at the local minimum $\epsilon = 0$, then $\bar V$ has an exact inflection point: $\partial_\sigma \bar V_c = 0$ and $\partial_\sigma^2 \bar V_c = 0$, regardless of the shape of the non-canonical kinetic term $\bar K$. Any potential with a near inflection point can always expressed as $\bar V = V_{0} + \delta V$, where $V_{0}$ has an exact inflection point and $\delta V$ is a small perturbation. Then, $\delta V$ will only influence the second phase of inflation, but not the first one that is tested by observations of the CMB. From Eq.~\eqref{def:slowroll_params}, it follows that $\epsilon_{c} \propto (\partial_\sigma\delta V_{c})^{2}$ and $\xi_{c} \propto \partial_\sigma\delta V_{c}$. Therefore, Eqs.~\eqref{eq:N2} and \eqref{eq:PRc} imply that \begin{equation}
	\prs(k_{c}) \propto N_{2}^{4},
\end{equation}
as we vary $\delta V$ while keeping $V_{0}$ constant. The overall proportionality factor depends on $V_{0}$.

\subsubsection{Beyond the slow-roll approximation}

Consider now the case in which the slow-roll approximation no longer holds, as Hubble friction cannot ensure $\epsilon_{V} \approx \epsilon_{H}$ at all times. In this case it is necessary to solve the exact equation \eqref{eq:exact_phiN} to correctly describe the transition period between the two phases of slow-roll inflation. There are two qualitatively different possibilities: 

\begin{itemize}[leftmargin=*]
\item[i)] \textbf{Approximate inflection point} ($\mathbf{\epsilon_{V} \gtrsim 0}$) \quad In this case there can be no singularities in the power spectrum. Nevertheless, if $\epsilon_{H} \gg \epsilon_{V}(\phi_{c})$, as the field is rolling over the near-inflection point, the peak \eqref{eq:PRkc} found in the slow-roll approximation diminishes and may even vanish completely. This effect is generally accompanied by a significant reduction of the length of the second inflationary phase as the field rolls over the inflection point without being slowed down considerably. It is then important to calculate the evolution of the field exactly by solving Eq.~\eqref{eq:exact_phiN}, rather than relying on the slow-roll approximation which here delivers inaccurate results.  Most of the models of single field double inflation considered in the literature~\citep{Bullock:1996at,Garcia-Bellido:2017mdw,Ezquiaga:2017fvi} belong to this class, however, they only consider the slow-roll approximation. An interesting extremal case is given by potentials that are flat  within a small field interval, e.g. an exact inflection point. This leads to ultra slow-roll inflation for which the slow-roll approximation inevitably fails~\cite{Tsamis:2003px}.

\item[ii)] \textbf{Local minimum} \quad The slow-roll approximation predicts that the field stops at the minimum, and is thus incapable of describing scenarios in which the inflaton possess sufficient kinetic energy to climb over the local potential barrier. According to the exact solution of the equation, however, the inflaton can exit the local minimum and give rise to a second phase of inflation. Then, as the field slow-rolls down from the local maximum $\phi_{c}$, the inflationary dynamics recovers hilltop inflation~\citep{Boubekeur:2005zm} with the potential
\begin{equation}\label{eq:HT_potential}
	V = V(\phi_{c}) \left[ 1 - \frac{\eta_{c}}{2}\left(\frac{\phi - \phi_{c}}{\mpl}\right)^{2} + \mathcal{O}(\phi - \phi_{c})^{3} \right],
\end{equation}
where $\eta_{c} \equiv |\eta_{V}(\phi_{c})|$. Notice that a few $e$-folds of inflation are produced even as the inflaton climbs uphill. It follows from Eq.~\eqref{eq:Ntot} that close to the maximum, when $\phi - \phi_{c} \ll \mpl \sqrt{2/\eta_{c}}$, the number of $e$-folds depends logarithmically on $\phi - \phi_{c}$. Therefore
\begin{equation}\label{eq:HT_eps}
	\epsilon_{V} \propto  e^{-2\eta_{c} N_{2}}.
\end{equation}
The slow-roll parameter $\epsilon_{V}$ is therefore exponentially suppressed in the beginning of the second phase and the corresponding curvature fluctuations obey a power law with spectral index 
\begin{equation}
	n_{s2} \approx 1 - 2\eta_{c}.
\end{equation}
Notice that the height of the peak in $\prs$ depends on how close to $\phi_{c}$ the field enters slow-roll. Moreover, the potential energy remains practically constant  during slow-roll, then \eqref{eq:PRk} implies that the height of the peak scales exponentially with the duration of the second phase 
\begin{equation}\label{eq:PRk_ii}
	\prs(k_{c}) \propto e^{2\eta_{c} N_{2}}.
\end{equation}
For a fixed $N_{2}$ it follows that a smaller spectral index generally yields a higher peak.  A single field double inflation model of this type has been considered in~\citep{Yokoyama:1998pt}.
\end{itemize}

\noindent  We find that models predicting a first phase of inflation consistent with observation and a second phase that creates a large peak in the curvature fluctuation power spectrum tend to fall in the second category.\footnote{These scenarios may also be modified by quantum effects such as tunneling. This may shorten the second phase even more. In the following we will, however, restrict our discussion to classical dynamics only.}

\section{Double inflation with running self-coupling} 
\label{sec:model}

Having the general results at hand, we now present a single field inflation model motivated by particle physics. We consider a general Jordan frame scalar potential
\begin{equation}
  V(\sigma) = \frac{1}{2} m^{2} (\sigma-v_{1})^{2} + \frac{1}{3} \mu (\sigma-v_{1})^{3} + \frac{1}{4} \lambda (\sigma-v_{1})^{4},
  \label{eq:V:J}
\end{equation}
with a canonical kinetic term for $\sigma$, corresponding to $K(\sigma) = 1$, and a non-minimal coupling
\begin{equation}
  \Omega(\sigma) = 1 + \frac{\xi}{\mpl^{2}} \sigma^{2}.
\end{equation}
In Eq.~\eqref{eq:V:J} we denote with $v_{1}$ the vacuum expectation value of the Jordan frame field $\sigma$. We find it convenient to express the Einstein frame potential  $\bar{V}$ in terms of its extremum solutions $v_{i}$. The extremum equation for $\bar{V}$ can be written in the general form%
\footnote{While it would seem that the numerator is a 5th order polynomial in $\sigma$, the leading term always cancels out, as can be easily seen.}
\begin{equation}\label{eq:extr:V:bar}
	0 = \partial_{\sigma} \bar{V} = -\frac{\xi \mu \, \mpl^{4}}{3(\mpl^{2} + \xi\, \sigma^{2})^{3}} \prod_{i = 1}^{4} (\sigma - v_{i}).
\end{equation}
The potential is then given by
\begin{equation}
	\bar{V} = \int_{v_{1}}^{\sigma} \partial_{\varsigma} \bar{V}(\varsigma) d\varsigma,
\label{eq:V:bar:int}
\end{equation}
where $\sigma = v_{1}$ is the global minimum and, by construction, $\bar{V}(v_{1}) = 0$. In general the integral of \eqref{eq:extr:V:bar} has a non-polynomial term, which will vanish provided that the extrema $v_{i}$ satisfy
\begin{equation}
	3\mpl^{4} + \mpl^{2} \xi\,e_{2}(v_{i}) + 3 \xi^{2} e_{4}(v_{i})  = 0,
\end{equation}
where $e_{2}$ and $e_{4}$ are elementary symmetric polynomials of $v_{i}$.  This constraint eliminates one parameter, $v_{4}$ for instance. Another parameter is, however, introduced as the constant of integration resulting from \eqref{eq:V:bar:int}, which we set by choosing of the lower limit. The case of a potential with an exact inflection point is achieved by setting $v_{2} = v_{3}$. A near inflection point corresponds to $v_{2}^{*} = v_{3}$ but $v_{2} \neq v_{3}$ and the potential has two local minima if $v_{2,3}$ are real but $v_{2} \neq v_{3}$.

\begin{figure}[t]
\begin{center}
  \includegraphics{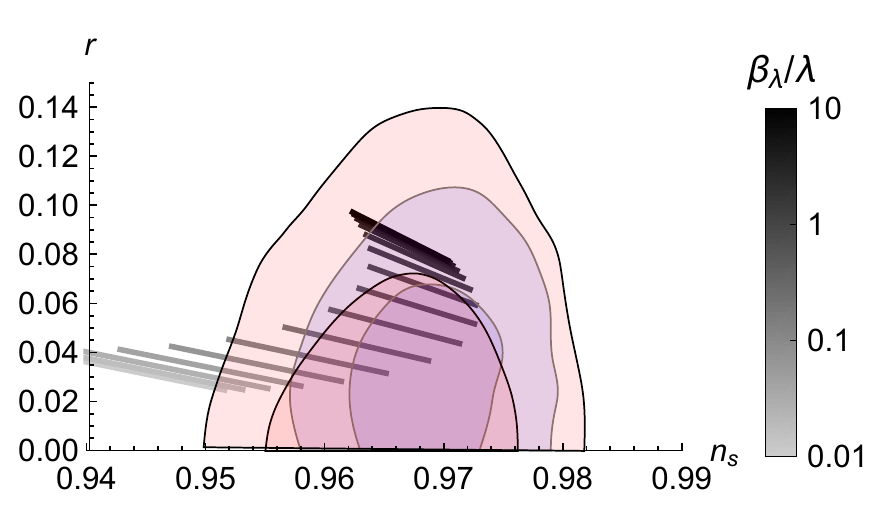}
\caption{The tensor-to-scalar ratio $r$ vs. the spectral tilt $n_{s}$ from $\beta_{\lambda}/\lambda = 0.01$ (bottommost line) to $\beta_{\lambda}/\lambda = 10$ (topmost line). 
We show the $1\sigma$ observational bounds \citep{Ade:2015lrj} for a running $n_{s}$ (black boundaries)  and for a  not running  $n_{s}$ (gray boundaries). 
 We choose $v_2 \approx v_3 = 0.106$, $\xi = 122.7$, $m_{\Phi} = 0.12 \mpl $.}
\label{fig:r:ns}
\end{center}
\end{figure}

As discussed in section~\ref{sec:infphase2}, in the slow-roll approximation the field stops indefinitely at an exact inflection point. In our case, instead, due to the breakdown of slow-roll regime between the two stages, the field acquires enough kinetic energy to cross such a point. By studying numerical solutions of the exact field equations \eqref{eq:exact_phiN} we find that a viable number of $e$-folds accompanied by a large peak in the power spectrum can not be achieved if $\bar V$ has a (near) inflection point. Instead, these features are only present in scenarios where the potential has a shallow minimum at $v_{3}$ and a maximum at $v_{2}$. The inflaton rolls into the minimum, slows down and enters the second stage of slow-roll inflation shortly after crossing the maximum.

We choose the parameters in a way that the scalar self-coupling $\lambda$ is small in the energy range where the second stage of inflation takes place, so that the parameters of the potential do not run much. In order to fit the inflationary measurements, however, we introduce another scalar field $\Phi$, with a mass $m_{\Phi}$, which does not directly participate in the inflation but which  couples to the inflaton via a quartic term $\lambda_{\phi\Phi} \phi^{2} \Phi^{2}$, generating the $\beta$-function $\beta_{\lambda} \propto \lambda_{\phi\Phi}^{2}$. Thus, we parametrise the renormalisation group (RG) running of $\lambda$ as
\begin{equation}
  \lambda(\sigma) = \lambda({m_{\Phi}}) + \theta(\sigma - m_{\Phi}) \beta_{\lambda} \ln \frac{\sigma}{m_{\Phi}},
\end{equation}
where $\theta$ is the unit step function. The running of the inflaton self-coupling  makes the potential steeper and, somehow counter-intuitively,  allows the inflation to begin at a straighter slope which generates the inflationary parameters $n_{s}$ and $r$ in the correct range.\footnote{The running will generally modify the positions of the extrema found by \eqref{eq:extr:V:bar}. Thus, for simplicity, we will consider the case where $m_{\Phi} \geq v_{2}$. In that case the first phase will be described by a scenario with a running coupling and the second phase corresponds to a hilltop like inflation.}

We plot in Fig.~\ref{fig:r:ns} the model predictions for the tensor-to-scalar ratio $r$ as a function of the spectral index $n_{s}$ for the parameters specified in the caption, which comply with all the CMB constraints despite the relatively small number of $e$-folds during the first phase of inflation. Notice that the inflationary parameters $r$ and $n_{s}$ depend on the ratio $\beta_{\lambda}/\lambda$ only, while the overall normalisation of the potential is fixed by the amplitude of scalar perturbations.  The results (grey lines) are for $25\div 35$ $e$-folds during the first stage of inflation. The observables depend on the value of the $\beta$-function that is varied from $\beta_{\lambda}/\lambda = 10^{-2}$ (bottommost line) to $\beta_{\lambda}/\lambda = 10$ (topmost line) 
in steps of $1/10$, as described by the colour code in the figure.  
The $1\sigma$ observational bounds \citep{Ade:2015lrj} are given for the cases of  running $n_{s}$ (black boundaries) and not running $n_{s}$ (grey boundaries). 
We therefore find that our model can easily predict $(n_s,r)$ within the $1\sigma$ confidence range of the present measurements.

\begin{table*}[t]
\begin{center}
\begin{tabular}{ccccccccccc}
$v_{2}/\mpl$ & $v_{3}/\mpl$ & $\xi$ & $\lambda/10^{-5}$ & $\beta_{\lambda}/\lambda$ & $m_{\Phi}/\mpl$ & $n_{s}$ & $r$ & $\prs(N_{\rm max})$ & $N_{\rm max}$
\vspace{1mm}
\\
\hline 
$0.09304$ & $0.10328$ &$86.449$ & $3.17$ & $0.3$ & $0.11$ & $0.960$ & $0.060$ & $9.6\times 10^{-4}$ & $38$
\\
$0.09303$ & $0.10328$ &$86.450$ & $3.58$ & $0.3$ & $0.11$ & $0.956$ & $0.066$ & $4.9 \times 10^{-3}$ & $35$
\\
$0.18346$ & $0.20937$ &$21.741$ & $0.167$ & $0.35$ & $0.22$ & $0.966$ & $0.056$ & $3.8\times 10^{-4}$ & $42$
\\
$0.02080$ & $v_{3}=v_{2}$ &$2944.0$ & $3460$ & $0.42$ & $0.022$ & $0.962$ & $0.067$ & $7.2\times 10^{-7}$ & $39$
\end{tabular}
\end{center}
\caption{The model parameters for the chosen benchmark points. In all cases $v_{1} = 0$ and $N_{1} + N_{2} = 60$. $\prs(N_{\rm max})$ denotes the maximal amplitude of the scalar perturbations and $N_{\rm max}$ is the time in $e$-folds when the maximum is reached. $v_{3}=v_{2}$ indicates an exact inflection point.}
\label{tab:bench}
\end{table*}

\begin{figure*}[tb]
\begin{center}
  \includegraphics[width=0.46\textwidth]{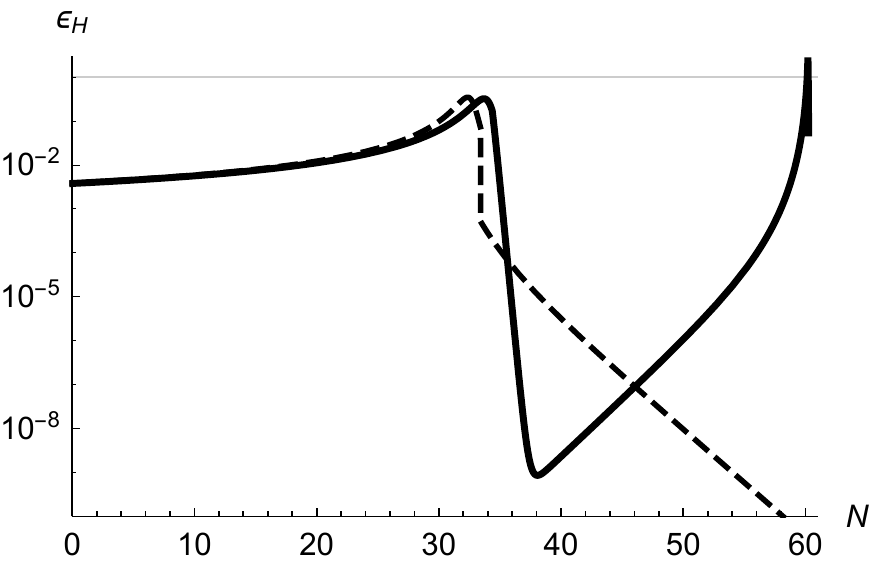}~\includegraphics[width=0.51\textwidth]{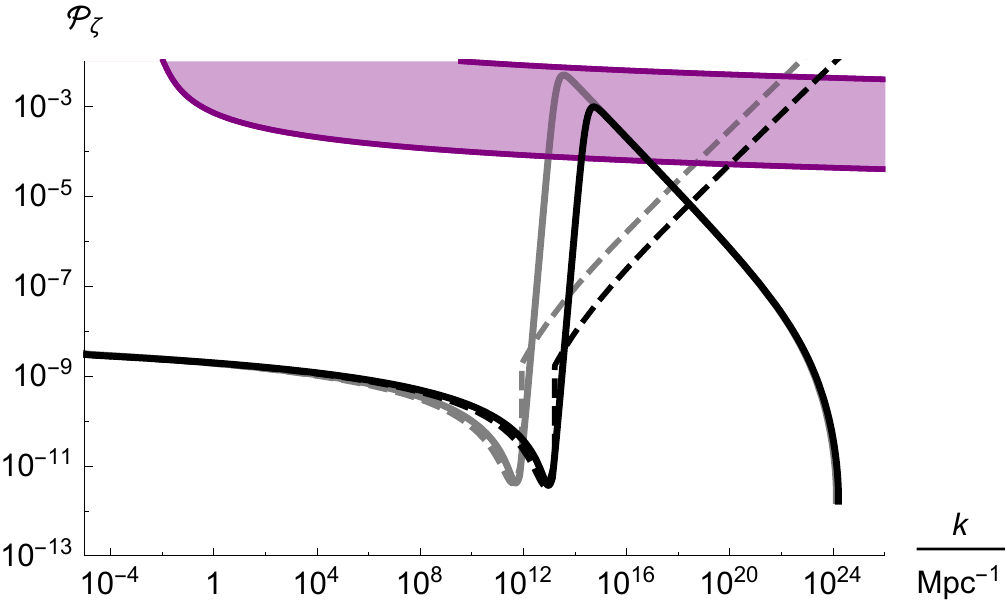}
\caption{Left: The Hubble slow-roll parameter $\epsilon_{H}$ obtained with the exact approach (solid line) and using the slow-roll approximation (dashed line) as functions of the number of $e$-folds $N$. Right: The power spectrum $\prs$ as a function of the wave number $k.$ The potential parameters are given by the first (black lines) and the second (gray lines) benchmark point in Table~\ref{tab:bench}. The purple band highlights the  curvature fluctuations yielding a PBH population that, depending on the value of the criticality parameter $\zeta_c$, can match the measured DM abundance. }
\label{fig:eps}
\end{center}
\end{figure*}

We present the parameters  of our model for a few chosen benchmark points in Table~\ref{tab:bench}. We always take the global minimum at $v_{1} = 0$. Notice that $\beta_{\lambda}$ is fairly small,  thus $\lambda_{\phi_{\Phi}}$ does not spoil the perturbativity of the model.  To analyse the dynamics of inflation in this model we plot in Fig.~\ref{fig:eps} the behaviour of the relevant inflationary parameters for the first benchmark point in Table~\ref{tab:bench}.  The left panel of Fig.~\ref{fig:eps} shows the Hubble slow-roll parameter $\epsilon_{H}$ as a function of the number of $e$-folds $N$. The dashed line corresponds to the solution obtained assuming the slow-roll approximation, whereas the solid line shows the exact solution. The thin horizontal line, $\epsilon_{H} = 1$, indicates the end of inflation. In the case of the first benchmark model the Universe is inflating even between the two phases, but the slow-roll approximation fails to describe the transition.  

The right panel shows the resulting power spectrum $\prs$ as a function of the wave number $k$. We see that the effect of the second phase of inflation is to produce a large peak at large values of $k$, which lasts for roughly 24 $e$-folds. The power spectrum at the second stage is well approximated by a power law with a spectral index $n_{s} = 0.4$. The exponential dependence of $\epsilon_{H}$ on $N$ matches well the prediction \eqref{eq:HT_eps} of hilltop inflation. Comparison of the first two benchmark points from Table \ref{tab:bench} as depicted in the left panel of Fig.~\ref{fig:eps} shows the exponential dependence in Eq.~\eqref{eq:PRk_ii} of the height of the peak in the curvature spectrum.

\section{Primordial black holes from the single field double inflation} 
\label{sec:PBH}

As was demonstrated above, the single field double inflation scenario decouples the dynamics of large cosmological scales probed by the CMB experiments from that of smaller scales which affect, for instance, the structure formation. In this section we show by the precise computations that our model can  also explain the measured DM relic abundance by giving rise to a PBH population  originated by large curvature fluctuations on small cosmological scales. In our scheme, such perturbations exit the inflationary horizon during the earliest of the inflationary epochs and, as they re-enter, the energy density contained in the associated volume collapses to a black hole~\citep{Hawking:1971aa,Carr:1974nx,1975ApJ...201....1C,1975A&A....38....5M,CHAPLINE:1975aa}. The spectrum of the perturbations generated by the inflaton in this inflationary stage, therefore, determines the mass function of the PBHs.

The hypothesis that DM is constituted by a relic abundance of PBH is severely constrained by astrophysical and cosmological observations. The strictest bounds arise from the black hole evaporation~\citep{Page:1976wx,Sreekumar:1997un,Carr:2009jm}, femtolensing~\citep{Barnacka:2012bm}, microlensing~\citep{Allsman:2000kg,Tisserand:2006zx,Griest:2013aaa,Niikura:2017zjd}, energy injection during the CMB era by black hole accretion~\citep{Ricotti:2007au,Clesse:2016ajp,Ali-Haimoud:2016mbv,Blum:2016cjs}, neutron star capture~\citep{Capela:2013yf}, white dwarf explosions~\citep{Graham:2015apa}, survival of stars in dwarf galaxies~\citep{Brandt:2016aco,Koushiappas:2017chw}, distribution of wide star binaries~\citep{Monroy-Rodriguez:2014ula} etc. (for a detailed description see~\citep{Carr:2009jm,Carr:2016drx,Carr:2017jsz}). We remark, however, that many of these constraints rely on different assumptions and involve uncertainties, in a way that not all of them should be taken on equal footing. In addition to that, the mentioned constraints might also depend on the mass function of the PBHs, which is therefore a key quantity in assessing the viability of this picture~\citep{Carr:2017jsz}.

The proposed scenario enforces on small cosmological scales large curvature fluctuations, $\zeta$, that result in a PBH population possibly characterised by a broad mass spectrum. In more detail, the fractional contribution of a PBH population of mass $M_{\rm PBH}$ to the energy content of the Universe at the formation time $t_M$ is quantified by~\citep{Harada:2013epa}
\begin{equation}\label{eq:betaBH}
	\beta_{0}(M_{\rm PBH}(N))
	\equiv\frac{\rpbh(M)}{\rtot}\Bigg|_{t=t_M}
	\approx \sqrt{\frac{2\prs}{\pi \zeta_{c}^2}} \exp\left(-\frac{\zeta_{c}^2}{2\prs}\right),
\end{equation} 
where we have assumed that the distribution of primordial density fluctuation, as well as the proportional curvature perturbations, be Gaussian, and that $\zeta_c^{2} \gg \prs$. It is then clear that peaks in the curvature power spectrum regulate with their widths the mass spread of the resulting PBH population. The value of the parameter $\zeta_c$ is the topic of several dedicated studies~\citep{Garcia-Bellido:2017mdw,Clesse:2015wea,Harada:2013epa} which confine it to the interval $(0.07, 0.7)$~\citep{Harada:2013epa}.

The mass of the PBH resulting from the collapse of curvature fluctuations during the radiation dominated era is of the same order of the horizon mass $M_H$: $M_{\rm PBH} \approx 0.2 M_{H}$ \citep{Carr:1975qj}. The latter is defined as the mass inside the horizon or, equivalently, the mass of a black hole whose Schwarzschild radius matches the Hubble radius
\begin{equation}\label{eq:MH}
	M_{H} = \frac{4 \pi}{3H^{3}}\rho = \frac{4 \pi \mpl^2}{H}.
\end{equation}
Since its formation, the fraction of PBH \eqref{eq:betaBH} scales  as $a^{-3}$, in the same fashion as cold matter. As the total energy density of the Universe scales instead as $a^{-4}$,  the relative contribution of PBH to the total energy density grows as $a$ until matter-radiation equality. From Eq.~\eqref{eq:MH} it follows that the horizon mass scales as $M_{H} \propto a^{2}$ during radiation domination. In conclusion, the PBH abundance at the time of matter-radiation equality is then given by
\begin{equation}
	\Omega_{\rm PBH}^{\rm eq} 
	=  \int_{M^*}^{M^{\rm eq}} \frac{\td M}{M}\, \beta_{\rm eq}(M),	
\end{equation}
where $M_{\rm eq} = 3.2 \times 10^{17} \msol \, \left(\Omega_{m}/0.31\right)^{-2}$ is the horizon mass~\citep{Green:2004wb} and
\begin{equation}
	\beta_{\rm eq}(M) = \sqrt{\frac{M_{\rm eq}}{M}} \beta_{0}(M)
\end{equation}
the mass function at matter-radiation equality. To relate the horizon mass to the wavenumber at the horizon entry, $k = aH$, notice that during the radiation dominated epoch $k \propto a^{-1} \propto M_{H}^{-1/2}$, implying that 
\begin{equation}
	M_{H} 
	= M_{\rm eq} \left(\frac{k}{k_{\rm eq}}\right)^{-2}
	=  3.2 \times 10^{13} \msol 
	\left(\frac{k}{\Mpc^{-1}}\right)^{-2}\,,
\end{equation}
where $k_{\rm eq} = 0.01 \left(\Omega_{m}/0.31\right)\Mpc^{-1}$  is the wavenumber of the perturbation entering the horizon at matter-radiation equality~\citep{Green:2004wb}. 

The largest theoretical uncertainty in the computation comes from the parameter $\zeta_c$. In order to estimate whether the model we consider is able to yield a sizeable fraction of the measured DM  abundance, we show in the right panel of Fig.~\ref{fig:eps} the density perturbations $\prs$ corresponding to $\beta_{\rm eq} = 1$ for $\zeta_c$ in the conservative range $ (0.07, 0.7)$. These extremal cases correspond to
\begin{equation}
	\prs{}_{,c} \equiv \frac{\zeta_{c}^{2}}{W\left( \frac{2}{\pi} (k/k_{\rm eq})^{2}\right)} \,,	
\end{equation}
where $W$ denotes the Lambert $W$-function. As in the early Universe $k \gg k_{\rm eq}$ the condition $\zeta_{c}^{2} \gg \prs{}$ assumed in \eqref{eq:betaBH} is naturally satisfied in case the PBH are light. As shown in Fig.~\eqref{fig:eps}, the peak in the spectrum is high enough to produce an abundance of PBH that matches the measured DM one.

Regarding the shape of the mass function\footnote{In the scenario we consider, the PBH formation proceeds exclusively in the radiation-dominated era and concludes at the matter-radiation equality. The information on the generated PBH abundance is therefore encapsulated in  $\beta_{\rm eq}$.}. In the case where the slow roll approximation holds between the two phases of inflation, we find upon an expansion of the power spectrum in Eq.~\eqref{eq:PRkc} up to second order in $\ln M$ a lognormal shape that has been often considered in the literature~\citep{Dolgov:1992pu,Blinnikov:2016bxu,Green:2016xgy,Horowitz:2016lib,Kuhnel:2017pwq,Dolgov:2017nmh},
\begin{equation}
	\beta_{\rm eq}(M)
	\approx 	\frac{\Omega_{\rm PBH, eq}}{\sqrt{2\pi}\sigma}
			\exp\left( - \frac{\ln^{2}  (M/M_{c})}{2\sigma^{2}}  \right)\,,
\end{equation}
where 
\begin{equation}
	\sigma = \zeta_c^{-1} \sqrt{\frac{\prs{}_{\rm i}}{2\xi_{\rm i}}} , 
	\qquad
	\Omega_{\rm PBH, eq} = \sqrt{\frac{1}{2\xi_{\rm i} } \frac{M_{\rm eq}}{M_{c}}}  \frac{2\prs{}_{\rm i} }{\zeta_c^{2}} \exp\left( -\frac{\zeta_c^{2} }{2\prs{}_{\rm i} } \right)\,,
\end{equation}
denote the width of the mass function and the PBH abundance at matter-radiation equality. $\xi_{\rm i} $ and $\prs{}_{\rm i} $ are calculated at the near inflection point. 

The analytic treatment of the mass function is more involved in more realistic cases where the slow-roll condition is violated. Nevertheless, given that the power spectrum has a peak at $k_{c} \propto \exp(N_{c})$, then whenever the second order expansion
\begin{equation}
	\epsilon_H(N) = \epsilon_H(N_c) + \frac{\partial_{N}^{2}\epsilon_H(N_c)}{2}(N - N_c)^{2} + \mathcal{O}(N - N_c)^{3}
\end{equation}
holds during the full period of PBH production, the resulting mass function may be well approximated by a lognormal shape. As deviations from this expansion are negligible for shorter periods of PBH formation, it is expected that the lognormal approximation is more accurate for narrower mass functions. The second phase of hilltop inflation produces a power-law tail extending to smaller scales. Therefore, broad mass functions produced by single field double inflation tend to be skewed towards lighter masses.\footnote{This tail will not follow a power law since Eq.~\eqref{eq:betaBH} implies that power-law curvature fluctuations $\prs$ will not generate a power-law tail for the mass function.}

\section{Conclusions} 
\label{sec:Conclusions}

We presented a study of single field double inflation in the context of general scalar-tensor theories and present the general qualitative features of these scenarios based on the \emph{exact} classical dynamics of the scalar field. Requiring a UV complete scenario and a renormalisable inflaton potential, we proposed a model that gives rise to a first phase of inflation which complies with the measurements of the cosmic microwave background fluctuations and spans about $N_1={\cal O}(30)$ $e$-folds. Such an inflationary era is supported by a potential that involves both the inflaton non-minimal coupling to gravity and logarithmic corrections to its self-coupling due to the renormalisation effects. Owing to the latter, the predictions of our potential differ from those of Starobinsky inflation, in line with similar models based on Coleman-Weinberg type of potentials. 

The first phase of inflation, is followed by a second era of hilltop inflation which lasts for about $N_2={\cal O}(30)$ $e$-folds and is able to give rise to a sizeable PBH population. In general, we find that the slow-roll condition breaks down between the two inflationary stages, leading to inaccurate results. To overcome this problem, we computed the dynamics of the inflaton field exactly, going beyond the approximations generally adopted in literature.    

The model we consider matches the values of the spectral index $n_s$ currently favoured and predicts a tensor-to-scalar ratio $r$ of the order of 0.05,  potentially measurable in next-generation experiments. The power spectrum of curvature fluctuation is characterised by a peak at small scales, which allows PBHs to match the measured DM abundance. The mass function describing the PBH span is approximately lognormal.

\section*{Acknowledgements} 
\label{sec:Acknowledgements}
The authors thank Bernard Carr, Juan Garcia-Bellido, Antonio Racioppi, Ville Vaskonen and Cristiano Germani for numerous discussions on topics of this paper.
This work was supported by the ERC grants IUT23-6, PUT799, PUTJD110 and by EU through the ERDF Center of Excellence program grant TK133.

\bibliographystyle{JHEP}
\bibliography{citations}

\end{document}